\documentclass[12pt, letterpaper]{article}

\usepackage{amsmath,amssymb}
\usepackage{cite}
\usepackage{fancyhdr}
\usepackage[top=1in, bottom=1in, left=1in, right=1in]{geometry}
\usepackage{graphicx}
\usepackage{hyperref}

\numberwithin{equation}{section}
\date{}

\title{{\rm\footnotesize \qquad \qquad \qquad \qquad \qquad \ \qquad \qquad \qquad \ \ \ \ \ \                       RUNHETC-2015-27     
SCIPP 15/31}\vskip.5in    Current Algebra on the Conformal Boundary and the Variables of Quantum Gravity}
\author{Tom Banks\\
NHETC and Department of Physics \\
Rutgers University, Piscataway, NJ 08854-8019\\
{\it and}\\
Department of Physics and SCIPP\\
University of California, Santa Cruz, CA 95064\\
E-mail: \href{mailto:banks@physics.rutgers.edu}{banks@physics.rutgers.edu}
}
\begin{document}

\maketitle
\thispagestyle{fancy} 

\begin{abstract} I argue that scattering theory for massless particles in Minkowski space should be reformulated as a mapping between past and future representations of an algebra of densities on the conformal boundary.  These densities are best thought of as living on the momentum space light cone dual to null infinity, which describes the simultaneous eigenstates of the BMS generators. The currents describe the flow of other quantum numbers through the holographic screen at infinity.  They are operator valued measures on the momentum light cone, with non-zero support at $P = 0$, which is necessary to describe finite flows of total momentum, with zero energy-momentum density, on the asymptotic holographic screen.  Jet states, the closest approximation to
the conventional notion of asymptotic particle state, have finite momentum flowing out through spherical caps of finite opening angle, with the zero momentum currents vanishing in annuli surrounding these caps.  
Although these notions are valid both in field theory and quantum gravity, I'll argue that they form the basis for understanding the holographic/covariant entropy principle in the latter framework, where the densities form a complete set of operators. The variables on a finite area holographic screen are restrictions of those at infinity.  The restriction is implemented by a cutoff on the Euclidean Dirac spectrum on the screen, which is a generalized UV/IR correspondence.

\normalsize \noindent  \end{abstract}


\newpage
\tableofcontents
\vspace{1cm}


\section{Boundary Gauge Invariance}

The earliest indication that the quantum theory of gravity is holographic in nature comes from the Wheeler-deWitt equation
\begin{equation} \frac{\delta S}{\delta g^{00}} (x) = 0 , \end{equation} which says that the traditional definition of the energy density in quantum field theory, vanishes.  Indeed, in all cases where we understand it, the Hamiltonian for quantum gravity is a boundary integral.   The local energy density is a generator of gauge transformations, which should vanish on physical states. How then are we to understand local bulk physics?  The clues, I believe, are to be found by examining gauge invariances of a simpler type, coupled with a profound argument due to Jacobson\cite{ted}. 

Suppose we ask the simpler question about local physics in a non-gravitational gauge theory.  In this case of course, we have gauge invariant local operators, and an energy density built from them, but these do not exhaust the degrees of freedom in a bounded region of space-time.  The set of gauge transformations, modulo those which vanish at the boundary, generate physical degrees of freedom in that region.  If the boundary is finite, they can be understood as non-local gauge invariant operators (Wilson loops), which penetrate the boundary.

In non-gravitational quantum field theories\footnote{It is of course my basic contention that quantum gravity is not a quantum field theory, but I'm using this phrase for the benefit of those readers who do not yet agree with me.} the energy operator is constructed as an integral over local gauge invariant operators, and these boundary DOF carry energy that can be derived from the local energy density in a slightly larger region.  In any theory of quantum gravity, as a consequence of the WD equation, this is not the case, and the Hamiltonian for a finite region must be constructed solely from the boundary DOF.  This begins to look peculiar when we look at a slightly larger region.  The formerly gauge invariant DOF are now gauge modes and must be presumed to be embedded in some way in the boundary DOF on the larger boundary. We can begin to see the area law for entropy\cite{bhthjfsb} emerge from these primitive considerations, as well as the principle that operators localized within a causal diamond (in the sense that they are boundary gauge transforms of a smaller diamond) are a sub-algebra of the full operator algebra on the boundary.

Jacobson showed long ago\cite{ted} that the area law for the entropy of diamonds led to the Einstein equations:  they are the local version of the first law of thermodynamics, applied to the energy as measured by a local Rindler observer in the limit of infinite acceleration.   Actually local thermodynamics does not capture the cosmological constant, which should be viewed\cite{tbwf} as a boundary condition on the large area vs. large proper time asymptotics of a causal diamond, rather than a local energy density. 

In this paper, I will try to argue that these intuitions lead to a conjecture for the correct degrees of freedom of quantum gravity.  Let's begin by thinking about asymptotically flat space-time.  In this case, the full Hilbert space should be generated by some algebra of operators on the conformal boundary of space-time.  These have traditionally been taken to be creation and annihilation operators of particles, but in theories with massless particles there is a problem with this description of the Hilbert space.  Even in dimensions where there are no infrared divergences in amplitudes containing a finite number of particles in initial and final states, there are still amplitudes where both incoming and outgoing states contain an arbitrary number of massless particles with total energy below any given cutoff $\epsilon$.  Particles that are contained within spherical caps of finite opening angle around the direction of the total energy going off at some angle\footnote{In the context of perturbative QCD, these were called {\it jets} by Sterman and Weinberg\cite{sw} . } can be combined into a single {\it jet} of quantum numbers, but we must also account for very low energy radiation in arbitrary directions.   This suggests a description of the asymptotic Hilbert space in terms of an algebra of currents on the conformal boundary.  The prototypical example of such currents are the Bondi-Metzner-Sachs (BMS) generators of the ``asymptotic symmetries of asymptotically flat space-time".  This infinite set of commuting operators is acted on by the true asymptotic symmetry group, which is the Lorentz group $SO (1, d - 1)$, the conformal group of the holographic screen of the maximal causal diamond in asymptotically flat space-time.  The BMS generators are delta function measures on the holographic screen, so if we diagonalize them all, the screen itself is encoded on their spectrum.  This is the null cone $P^2 = 0$, where we make the arbitrary convention that outgoing momenta on future null infinity carry positive energy.  This does not quite cover everything, because future and past time-like infinity are not associated with a null momentum.  

The situation is clarified (as many things will be) by retreating to a finite causal diamond.  No matter how large it is, the asymptotic trajectory of a massive particle can, and typically will, pass through its null boundary.  The only difference between massive and massless particles is that the momentum for massive particles includes both momentum $p (1, \bf{\bf{\Omega}})$ directed out of the diamond, and momentum $q ( 1, - \bf{\bf{\Omega}} )$ directed along the null boundary of the diamond.  Here 
$\bf{\bf{\Omega}}$ is a unit $d - 1$ vector in $d$ dimensional asymptotically flat space.  The product $pq$ is proportional to the mass of the particle.  Our expectation from string theory is that the only stable massive particles are either BPS states, or particles carrying discrete symmetries, which in the weak string coupling limit, represent the K-theory of the vector bundles describing BPS D-branes\cite{Ktheory}.  The asymptotic operator algebra should have operators which carry all of these quantum numbers. The upshot of this paragraph is that if we are willing to carry along a double cover of the null cone, with the two copies related by spatial reflection, we should be able to describe both massive and massless particles (more properly jets) in terms of an algebra of charges on the null cone $P^2 = 0$.  

To describe finite regions in space-time, we have to make a gauge choice.  The natural regions to talk about are causal diamonds, because these have a clear quantum meaning, which can be viewed as the quantum gravity avatar of Einstein locality in quantum field theory.  A causal diamond contained in another one should be identified with a tensor factor in the operator algebra of the larger diamond.  The theory of tensor factorization in infinite dimensional operator algebras is fraught with mathematical subtlety. This is where the holographic principle comes to our rescue.  A variety of arguments, the most persuasive of which is Jacobson's derivation of Einstein's equations from the local thermodynamics of infinite temperature quantum systems obeying the BHtHFSB\cite{bhthjfsb} connection between area and entropy, suggest that a causal diamond with finite area holographic screen is described by a finite dimensional Hilbert space.  The definition of tensor factor then becomes straightforward, if we define the asymptotic Hilbert space as a limit of finite area diamonds associated with a time-like trajectory with proper time $T \rightarrow \pm\infty$.  Each smaller segment along the trajectory defines a smaller tensor factor, associated with the causal diamond of that sector.  This kind of nested tensor factorization is most efficiently described in terms of sub-factors of the current density operator algebra.   We'll see that the general prescription of the Holographic Space-time (HST) formalism in asymptotically flat space is to decompose the current density algebra into eigenfunctions of an $SO(d - 1)$ invariant differential operator on the holographic screen\cite{tbjk}.  This is the sub-group of $SO(1, d - 1)$ leaving invariant a particular time-like geodesic in Minkowski space.  Different geodesics will correspond to different decompositions.   

Returning to the boundary at infinity, we label the currents $J_i (P)$ by the eigenvalues of the BMS generators
we have $P = p (\pm 1, \pm {\bf \bf{\Omega}}) , $ where $p \geq 0$ and ${\bf \bf{\Omega}}$ is a unit $d - 1$ vector.  The first $\pm$ sign refers to the positive and negative branches of the null cone, which we associate with outgoing and incoming states respectively.  The second minus sign refers to the fact that at each outward pointing normal vector ${\bf \bf{\Omega}}$, we might have operators associated in the past (future) with both inward and outward (outward and inward) directed normals, if our model contains stable massive particles. 

The current algebra should be covariant under the $SO(1, d-1)$ isometry group of the null cone.  This simply means that the currents should transform as tensor or spinor fields on the cone. The classical avatars of the boundary current algebra can be found in the bulk description of massless particles in terms of gauge invariant field equations.   The right hand side of any such field equation is a current  $J^A $, a $d - 1$ form with values in the Lie algebra of the gauge group. The current satisfies the covariant conservation law \begin{equation} D^A_B J^B  = 0 , \end{equation} as a consequence of the Bianchi identities for the gauge field strength.  The operator $D_A^B = d \delta_A^B + {\cal A}_A^B$, where ${\cal A}$ is the one form gauge potential, and the product is the wedge product. Asymptotically, we can write the gauge potential as the sum of a background plus a small fluctuation.  If we solve the asymptotic field equations for the fluctuation in terms of its source, then the term in the covariant derivative proportional to the gauge potential can be rewritten entirely in terms of the gauge field.  The asymptotic covariant conservation law becomes an ordinary conservation law of a new current involving the ``physical" asymptotic gauge excitations.

For example, in Yang-Mills theory, the covariant conservation law is 
\begin{equation} 0 = d J^a + f^{abc} A^b \wedge J^c = d J^a + f^{abc} A^b \wedge d * F^c  = d(J^a + f^{abc} A^b \wedge * F ). \end{equation} The first equality uses the linearized equation of motion for $A^a$, which is valid asymptotically, and the second uses the fact that $F^a \wedge *F^b $ is symmetric under interchange of $a$ and $b$.  The extra term in the asymptotic conserved current is the non-abelian current of free gauge bosons.  In these equations, the field strength $F$ just has the linearized Maxwell form.

Similarly, in gravitational theories  the covariantly conserved $3$ form matter stress tensor $T^A$ ($A$ are components in a vierbein basis of the tangent space)
gives rise to asympotically conserved BMS currents
\begin{equation} P^A =  T^A + G^A\end{equation}
The gravitational BMS current $G^A$ comes from the term ($\omega$ is the spin connection) $ \omega^{AB} T^B $ in the covariant conservation law $d T + \omega^{AB} T^B = 0$ by substituting the asymptotic values of $\omega$ and $T$ computed from 
\begin{equation} d e^A + \omega^{AB} e^B = 0 , \end{equation}
and the asymptotic gravitational field equation, when the vielbein satisfies the asymptotic condition $e^A \rightarrow \eta^A + c_d M_P^{\frac{2 - d}{2}} h^A. $ Note that, as long as we preserve the zero torsion condition\footnote{Of course this gets modified in the presence of matter fields with spin, but it is replaced by a similar universal equation involving the spin.}, terms of higher dimension than the Einstein equations are asymptotically negligible.   This is a consequence of the fact that curvatures go to zero asymptotically.  We are, of course, assuming that the cosmological constant is zero, which is an asymptotic boundary condition.
The asymptotic value of $\omega$ is linear in derivatives of $h^A$ and the asymptotic value of $T^A$ is linear in $d \omega$ , so the asymptotic value of
the homogeneous term in the covariant conservation law is proportional to $\omega d\omega = - d \omega^2  $.   The resulting term $G^A$ is the asymptotic stress tensor of gravitons, when it is evaluated between states in Fock space.

Returning to the general case, denote the asymptotic current by $j$. If we look at the conservation law $d j = 0$ on the boundary of a large but finite causal diamond, the $d - 1$ form current breaks into a a pair of transverse $d - 2$ forms $j_{u,v}$ and a $d-3$ form $j_{uv}$.  
\begin{equation} j =  j_u du + j_v dv + j_{uv} dudv, \end{equation} and the conservation law becomes
\begin{equation} \partial_u j_v - \partial_v j_u + d j_{uv} = 0. \end{equation} 

Assume that the asymptotic connection between causal structure and conformal factor is like that in Minkowski space.  As the proper time goes to infinity the area of the holographic screen of the diamond also goes to infinity, smoothly. The standard boundary conditions on fields at ${\cal I}^{\pm}$ imply that the stationary current $j_{uv}$ on the transverse sphere, goes to zero.  If we have only massless particles in the theory, then $j_u$ goes to zero in the infinite past and $j_v$ goes to zero in the infinite future.  Thus, in both the past and the future, the current produces local densities on the sphere, which become independent of the null coordinate.  

Although I suspect that the boundary current algebra approach will be useful in massless quantum field theories with no dynamical gravitational field, we will concentrate on models for which one of the currents in the algebra comes from the stress tensor.  The resulting density operators on the spheres at $\pm \infty$ are the Bondi-Metzner-Sachs supertranslations $P^A (\bf{\Omega} )$.
They form an abelian sub-algebra of the current algebra.  The fact that \begin{equation} [P^A (\bf{\Omega} ) , P^{B} (\bf{\Omega}^{\prime} ) ] \propto \delta  (\bf{\Omega}  ,\bf{\Omega}^{\prime} ) , \end{equation} follows from the conservation laws $\partial_u P^A_- = \partial_v P^A_+ = 0$, plus the fact that at fixed $u$ or $v$, different points on the sphere are (infinitely) space-like separated. It is part of the requirement that the theory be local.  

The fact that the coefficient in front of the delta function vanishes follows from more subtle considerations. The conformal boundary of Minkowski space has $SO(1,d - 1)$ as its group of conformal isometries, and the current algebra should be covariant under this group.  A non-trivial right hand side of the commutator, would introduce an extra anti-symmetric tensor current\footnote{Of course, the usual identification of super-translations as generators of asymptotic diffeomorphisms automatically gives us commuting generators.}.  If the spin connection were a propagating field, this current would be the source of its field equation, but theories with such a propagating field do not have a unitary S-matrix. We will return to the spin connection below.

The BMS generators\footnote{Conventionally, the BMS algebra is the semi-direct product of the super-translations and the Lorentz group. In order to avoid the problem of referring to both super-translations and supersymmetry in the same sentence, I prefer to use the phrase BMS generators to denote the super-translations.} are operator valued delta function measures on the sphere.  Their eigenvalues are null vectors $P^2 = 0$. The generator concentrated at solid angle $\bf{\Omega} $ must transform as a null vector under the subgroup of the Lorentz group leaving  $\bf{\Omega} $ invariant.  This means that it must have the form
\begin{equation} P^A ( \bf{\Omega} ) = p (1, \pm  \bf{\Omega}  ) \delta  (\bf{\Omega} , \bf{\Omega^{\prime}}), \end{equation} where $p$ is a real number.  By convention, $\pm $ values of $p$ will occur only for the generators on ${\cal I}_{\pm}$, enforcing the positive energy condition on states.   For a system containing only massless particles only one of the two orientations of the spatial momenta would be allowed.  Keeping both inward and outward pointing  spatial momenta allows us to describing massive particles.

The upshot of all of this is to describe the conformal boundary of Minkowski space not as the singular Penrose diamond, but as the null cone $P^2 = 0$.  That is, we identify points on the sphere with the spectrum of the super-translations.  This is analogous to using longitudinal momentum instead of longitudinal position in null plane quantization.  In the present context it has a number of advantages.  The Lorentz group acts on the null cone in a transparent manner.  Questions about boundary conditions at spatial infinity get converted into questions about behavior at the singular point $P = 0$.  Finally, we can describe massive particles as states for which, at some point on the null cone, the actual momentum is a sum of the form $ \alpha_+ p (1, + \bf{\Omega} ) + \alpha_- p (1, - \bf{\Omega}) $, with the mass squared proportional to $p^2 \alpha_+ \alpha_- $.  In the conventional position space approach to the conformal boundary massive particles all go through the singular points at time-like infinity.   Note that, in terms of our general discussion of conserved currents above, what distinguishes massive particles from massless ones is that both components $j_{u,v}$ are present in the past as well as the future.  This is clearly a much more smooth limit of the description of the situation in finite diamonds, where generic time-like trajectories penetrate the null portion of the boundary, carrying currents along the boundary as well as out of it.

Our preliminary discussion of physical observables in a finite causal diamond suggests that the entire operator algebra of a theory of quantum gravity in asymptotically flat space, should be visible at null infinity.  Algebras associated with finite causal diamonds should be sub-algebras of the asymptotic algebra.
We can classify additional generators of the algebra according to their commutation relations with the BMS super-translations $P^A $.  In dimension higher than $4$ we expect that the Lorentz generators are the only operators which don't commute with $P^A$\footnote{In four dimensions, we also have the Virasoro generators\cite{virasorotbothers}.  I do not believe that there is a Poincare invariant scattering theory for quantum gravity in less than four dimensions.}.  They have the form
\begin{equation} L_{AB} = P_A \frac{\partial}{\partial P_B} - P_B \frac{\partial}{\partial P_A}
+ \Sigma_{AB} (P)  , \end{equation} where $\Sigma (P)$ commutes with $P^A$ and is therefore a function of its eigenvalue.
In effective field theory, these generators should arise, as above, from the asymptotic form of the covariant conservation law
\begin{equation} D_{\mu} \frac{\delta S}{\delta \omega_{\mu} } = 0  . \end{equation} Here $\omega_{\mu} dx^{\mu} $ is the spin connection one form.  The Einstein action, in first order formalism is 
\begin{equation} S_E = \int A_{AB} (d \omega_{AB} + (\omega \wedge \omega )_{AB} ) , \end{equation} where $A_{AB}  = \epsilon_{AB a \ldots b} e^a \wedge \ldots \wedge e^b , $.  In $d$ dimensions $A_{AB}$ is a wedge product of $d - 2$ vielbein one forms.  

With the usual Einstein action, the field equations for $\omega$ are algebraic, and solving them evaluates it as the usual Christoffel connection plus a variation of the matter action w.r.t. $\omega$.
When this expression is evaluated, in ordinary field theory, in terms of asymptotic fields on the conformal boundary, it gives rise to Lorentz generators of the form above, with $\Sigma_{AB} (P)$ written as a sum of spin operators of all particles in the model.  

Other Yang-Mills fields will give rise to generators $J_a (P)$ transforming in the adjoint representation of the gauge group. If the gauge group has a center, we must decide whether the model has angularly localized excitations transforming under the center.  If it does, the asymptotic operator algebra should have generators transforming under the center.  We can ask the same question for the group $Spin (1,d - 1)$, which, by the spin-statistics theorem, is the same question as whether there are localized excitations on the null cone, which transform under the $Z_2$ gauge symmetry that defines Fermi statistics. These operators will anti-commute at unequal $P$.  Since the quantum numbers that characterize operators transforming under the center of the group are discrete, they do not have to appear in field theory as the asymptotic right hand side of a bulk field equation for a dynamical gauge field.
 
 Operators transforming under the discrete gauge groups of the theory will have the general form  $Q_{\alpha}^i (P)$, where $\alpha$ is a spinor index.
 They should anti-commute at unequal angles, since such points are at infinite space-like separation.     The most general algebra consistent with these
 conditions and covariant under $SO(1,d-1)$ is
 \begin{equation} [\bar{Q}_{\alpha}^i (P) , Q_{\beta}^j (Q) ]_+ = Z^{ij} \delta (P\cdot Q) M_{\mu} (P, Q) \gamma^{\mu}_{\alpha\beta} . \end{equation}  We have omitted terms proportional to higher rank $SO(1,d-1)$ tensors, since these would represent charges of infinite branes, and are not appropriate for describing particles going out through particular directions at null infinity.  We have also omitted terms transforming as scalars.  These are related to massive particles.  The algebra above is compatible with the condition
 \begin{equation} P_{\mu} \gamma^{\mu}_{\alpha\beta} Q_{\beta} (P) = 0. \end{equation}   This is the Cartan-Penrose condition, which identifies the generators as {\it pure spinors} associated with the null vector $P$.  In more down to earth language, it says that they are spinors in the space-like plane transverse to $P$ - elements of the spinor bundle over the spherical holographic screen.
 
 The past and future, $P^0 <$ or $> 0$, parts of the null cone refer to the past and future boundaries of null infinity, and they are tied together by the behavior of the operator $Q_{\alpha} (P=0)$ which is a non-trivial {\it half-measure} on the sphere, whose support in a given state tells us about soft particles flowing out to infinity in various directions.  Strominger\cite{strom} pointed out that this represented spontaneous breakdown of the BMS super-translation algebra, and was connected to the boundary conditions at spatial infinity, which Christodoulo and Klainermann\cite{christklein} showed were essential to the definition of classical gravitational scattering theory. We have implicitly made a choice of spatial orientation for the null vector $P$ at a fixed angle: outward pointing for $P_0 > 0$ and inward for $P_0 < 0$, as is appropriate for massless particles.  $M_{\mu} (P,Q)$ is a four vector function of the parallel null vectors $P$ and $Q$.  We choose it to be the smaller of the two, so that the zero momentum generators anti-commute with all of the others.  $M$ then stands for minimum.
 
 The labels $i,j$ should be thought of as labeling a basis of eigenfunctions of the Dirac equation on a compact manifold, with a cutoff that encodes the volume of the manifold in Planck units, and some information about the 
shape of the manifold, as sketched in \cite{tbjk}\footnote{The part of HST that deals with compact dimensions remains massively underdeveloped, and I'd like to suggest that it is a fertile field for progress in the subject.}.  The $Z^{ij}$ are differential forms on the (fuzzy) compact manifold, and should be viewed as the non-perturbative generalization of wrapped brane charges in perturbative string theory.

One of the things that people used to perturbative string theory have a hard time understanding is that, in general, wrapped brane charges for genuine particle states are bounded operators.  In string perturbation theory, we calculate the spectrum at $g_S = 0$, where the string length is infinitely greater than $ L_P$. 
In this limit, arbitrarily large wrapped brane charges correspond to states lighter than the Planck scale, where the distinction between black holes and particle states is not relevant.  For any finite value of $g_S$ there will be a maximum charge above which the ``particle" label is no longer appropriate. Particle/jet states are defined as those for which
\begin{equation} \int d^d P \delta (P^2) Q_{\alpha} (P) f^{\alpha} (P) , \end{equation} with the support of $f^{\alpha} (P)$ having arbitrarily small opening angle on the sphere makes sense.  On the other hand\cite{susskindhologram} black holes subtend a finite area region on the screen at infinity.  The significance of the bound on brane charges will be enhanced when, in the next section, we discuss subalgebras of the current algebra appropriate to the description of finite area causal diamonds.  The holographic bound on the dimension of the Hilbert space of a finite diamond does not allow an algebra with infinite dimensional representations.  For the spherically symmetric part of the algebra, we impose finite dimensionality with a cutoff on the Dirac spectrum (equivalently, the angular momentum).  This leads to a UV/IR correspondence generalizing that of AdS/CFT: diamonds of smaller size have fuzzier angular localization.  A similar cutoff on the spectrum of the internal Dirac operator on a torus, cuts off the spectrum of BPS charges associated with toroidal momentum and the holographic principle tells us that the same must be true for charges associated with wrapped branes.  The internal cutoffs are independent of the size of the holographic sphere in Planck units and so survive in the limit of infinite diamonds.  

The complete super-algebra describing all particle states in the model must have a finite dimensional unitary representation for fixed $P$.  The classification of all such algebras, containing one graviton and no massless particles of higher spin, has not been carried out.  It is the analog, in this way of thinking about quantum gravity, of the tree level study of consistent compactifications of string theory.  Like the latter project, we do not expect it to give an answer to the question of whether models of quantum gravity in asymptotically flat space {\it must} be supersymmetric.  The evidence from string theory that this is the case comes from the study of interactions.  The above algebra will contain the algebra of supersymmetry if, for one or more values of the internal indices $i$ and $j$, $Z_{ij} = 1$.  
A related question is whether one must require that the asymptotic algebra contain, for fixed $P$, some operator carrying spin, which {\it is} the limit of a bulk current operator that couples to a massless particle.  If that is the case, then the Coleman-Mandula\cite{cm}-Haag-Lopuzanski-Sohnius\cite{hls} theorems imply that the model is supersymmetric.  For the remainder of this paper, we will assume that supersymmetry is a requirement.

The null cone $P^2 = 0$ is not a manifold, and the generators $Q_{\alpha}^i (P)$ are half-measures with support at the singular point $P=0$.  The fundamental requirement of a quantum theory of gravity is that there is a unitary operator $S$, which maps the past generators, with $P_0 < 0$ into those in the future
\begin{equation} Q_{\alpha\ +}^i (P) = S Q_{\alpha}^i (P) S^{\dagger}  . \end{equation}
Note that Strominger\cite{strom} derives equations like this for the operators $P^A$ as conservation laws of spontaneously broken symmetries, whereas for our purposes they are more analogous to the relation between in and out creation and annihilation operators. We will comment on Strominger's interpretation in the conclusions. $P^A$ is, in our formalism, just a variable parametrizing the null boundary.   The actual momentum of states in a jet representation of the algebra depends on the choice of the spherical caps for which $Q(f) | jet \rangle \neq 0$.
Conservation of momentum in scattering is an additional constraint on the $S$ matrix.  As we'll see below, it's quite easy to build models in which the energy and angular momentum in the Hilbert space of a given time-like geodesic, are conserved.  Lorentz boost invariance is more difficult, and has not yet been implemented.
\subsection{The Definition of Jet States}

Jet states are defined as states annihilated by the smeared operators $\int Q_{\alpha\ \pm}^i (P) f^{\alpha\ \pm}_i (P) $ unless the half-measurable smearing functions $f$ satisfy the following criteria

\begin{itemize}
\item If $P$ is not at the tip of the null cone, the support of $f(P)$ is confined to a finite number of non-overlapping spherical caps.

\item The support of $f$ {\it at} the tip of the null cone (which is a $d - 2$ sphere), vanishes in $d - 2$ dimensional annuli surrounding each of the spherical caps.

\end{itemize}

In conventional particle physics language the meaning of these statements is that all massless particles of vanishing asymptotic energy density, which are close to a finite momentum jet, are included in that jet, while particles below some total energy cutoff are allowed to be emitted in arbitrary directions isolated from the jet.
A careful formulation of such jet isolation criteria in quantum field theory requires us to think about a total energy cutoff for the unmeasured IR gauge bosons which is finite but arbitrary.  This corresponds to thinking about a delta function measure localized at $P = 0$ in terms of a sequence of characteristic function measures localized near that point. In real experiments, one has a finite holographic screen and a finite low energy cutoff\footnote{Of course, the Sterman-Weinberg definition of IR safe cross sections has proven too unwieldy to use in real experiments.}.
In a quantum theory of gravity, finite holoscreens pose new conceptual problems, which we have adumbrated above.  In the next section we will outline the way they are treated in the HST formalism.

\subsection{Executive Summary}

Gauge theories in a finite causal diamond always have extra observables whose support is limited to the boundary of the diamond.  Theories invariant under diffeomorphisms are special because these are the {\it only} gauge invariant observables in the diamond.  It follows immediately that the gauge invariant observables in a given diamond, are a sub-algebra of the observables on the boundary of a larger diamond, which contains it.  This is part of the general statement of the Holographic Principle.  The full algebra of observables in a given causal patch of space-time, lives on the boundary of the maximal diamond available in the space-time.  The Holographic Principle also contains the statement that a complete quantum model of that space-time can be formulated entirely in terms of this maximal algebra.  In general, the identification of a particular causal diamond algebra as a factor in the maximal algebra is ``gauge dependent".  We'll see below that a convenient gauge choice is equivalent to the choice of a particular time-like trajectory in the space-time.

The minimal operator content of the asymptotic algebra comes from the asymptotic values of covariantly conserved currents, which appear as the source terms in the field equations for gauge fields.  Given the boundary conditions describing asymptotically flat space, these are always related to closed $ d - 1$ forms, which include contributions from the relevant gauge field at infinity.  If the associated current flows are related to massless particles, then there is only one surviving density component on each branch of null infinity.  Among these conserved currents, we always have the stress tensor, and the corresponding asymptotic densities are the commuting super-translations of the BMS algebra.
These densities have pointlike support on the sphere at infinity, and we've argued that they should be used to parametrize the conformal boundary of asymptotically flat space-time.  The null cone $P^2 = 0$ describes momentum flow into and out of the conformal boundary.  Flows containing massive particles are described by currents that have both incoming and outgoing components on both the past and future boundaries. 

The rest of the asymptotic operator algebra consists, apart from the orbital part of the Lorentz generators, of operators that commute with the BMS generators and can be thought of as operator valued measures (and half measures) on the null cone.  

\section{Finite Causal Diamonds}

As indicated above, I believe that much of the structure discussed so far is applicable to quantum field theories, which contain massless particles.  The true novelty of the quantum theory of gravity is the holographic principle.  In QFT, the algebra of asymptotic operators is a sub-algebra of the algebra of bulk local operators.  In models with a mass gap, this statement is encoded in the LSZ formulae, and the well known ambiguities in the local field algebra corresponding to a given S-matrix.  The holographic principle tells us that in models of quantum gravity, the reverse is true: the algebra of operators localized in a given causal diamond, whose holographic screen has finite area, is a finite dimensional sub-algebra of the algebra of asymptotic operators. We've given the intuitive arguments for this above.

The additional challenge of a quantum theory of gravity is to describe these local sub-algebras in a manner which depends as little as possible on a choice of coordinates on space-time. Local time evolution presents an additional problem.  The Wheeler-DeWitt equation is the statement that a local change in the choice of time coordinate should leave physics unaffected.

The formalism of Holographic Space-time (HST) attempts to resolve both these problems by referring to the definition of causal diamonds in terms of time-like trajectories.  A given causal diamond in space-time is associated with a finite proper time interval along some time-like trajectory, which goes from the infinite past to the infinite future\footnote{In cosmology, the time-line is only half-infinite.  The HST formalism has not been generalized to describe cosmologies with both a Big Bang and a Big Crunch. Perhaps it shouldn't be. Such a universe would have a finite dimensional Hilbert space, like an asymptotically dS cosmology.  However, unlike that case, it's likely to have a time dependent Hamiltonian\cite{tbwfcrunch}, and an evolution operator which scrambles the system more and more rapidly as the proper time approaches the Big Crunch singularity. Such a model would have only a very small set of observables, whose nature was essentially thermodynamic. }.  A given time-like trajectory can be viewed as a nested sequence of causal diamonds.  Consider two time-like trajectories, with some synchronization of their proper times. As an example, in an approximately FRW Big Bang cosmology, we can consider time-like geodesics and measure proper time from the Big Bang hyper-surface.  At any time, the causal diamonds along the two trajectories will have some intersection, and there will be a maximal area causal diamond contained in that intersection.  The intuitive idea of ``relativity", namely that there is a simple mapping of the observations made by detectors following the two trajectories, can be translated simply into quantum language.  Each trajectory corresponds to a quantum system with time dependent Hamiltonian $H(t, {\bf x})$, where ${\bf x}$ labels the trajectory.  The Hamiltonian operates in a Hilbert space whose dimension is determined by the maximal area causal diamond, encountered at infinite proper time.  However, the Hamiltonian splits into $H_{in} (t , {\bf x}) + H_{out} (t, {\bf x}) $, where, for a finite area diamond, $H_{in}$ acts only on a finite dimensional tensor factor of the full Hilbert space, and $H_{out}$ acts on its tensor complement.  Note that for space-times which are asymptotically flat, or FRW with zero c.c., the entropy of the Hilbert space ${\cal H}_{in} (t)$ scales like $t^{d-2}$ as $|t| \rightarrow \infty$.  

The HST formalism describes this Hilbert space in terms of operators which belong to the ``fuzzy spinor bundle over the holographic screen"\cite{tbjk}.   These are operators 
$\psi_{l_1 , l_2} $, where the labels stand for the spectrum of the Dirac operator on the holographic screen $S^{d-2} \times {\cal K}$.  ${\cal K}$ is a compact manifold of fixed finite volume in Planck units.  The finite volume of the holoscreen is implemented by a cutoff on the spectra of the Dirac operators $D_{S}$ and $D_{{\cal K}}$, plus the requirement that the algebra of the resulting finite set of operators has a finite dimensional unitary representation.  For the sphere, a simple eigenvalue cutoff is equivalent to an angular momentum cutoff, and the commutation relations are completely determined by rotational invariance.   For ${\cal K}$ the detailed nature of the cutoff {\it and} the operator algebra, encode features of the geometry of the manifold.  I will not go into details on this correspondence, since it is only partially worked out.  Instead, I'll assume that ${\cal K} $ is a torus of minimal radius in Planck units\footnote{Readers aware of the literature on string dualities may balk at this phrase, since many of those dualities appear to refer to manifolds much smaller than the Planck scale.  The HST resolution of this ``paradox" has to do with the fact that wrapping numbers on fuzzy manifolds are not quantized in the same way as those on continuous manifolds.}.   The full set of commutation relations is then that of a collection of fermions.

In $d - 2$ dimensions, the eigenspinors of the spherical Dirac equation, with eigenvalue cutoff of order $N$ can be arrayed as
$$\psi_{a k}^{s_1 \ldots s_{d-2} } , $$ where $1 \leq s_i \leq N$ and the expression is totally anti-symmetric in the $s_i$.  $a$ is a minimal $d - 2$ spinor index and $k$ labels the cutoff eigenspinors of the internal Dirac operator.  For the minimal torus compactification, the variables are real and the product of the number of $a$ and $k$ values is $32$.  The entropy of the Hilbert space scales like $N^{d-2}$ so $N \propto t$ for large $N$.  Note that if we make the number of internal eigenspinors larger by a factor of $\lambda$, the entropy scales like $\lambda$.  For a smooth internal manifold of dimension $D$, the number of eigenspinors will increase by a factor $\lambda$ if we increase the Dirac eigenvalue cutoff by a factor $\lambda^{1/D}$ .  Thus, we get the expected Kaluza-Klein scaling of entropy under a rescaling of lengths of the internal manifold if we identify the eigenvalue cutoff with $V^{1/D}/L_P$, where $V$ is the volume of a roughly isotropic $D$ manifold.  Thus we see a UV/IR connection both for the internal and the Minkowski dimensions.  

Another general feature of this HST approach to compactification is that moduli of the internal manifold are discrete.  This is a feature whose origin is the finite entropy implicit in the Holographic Principle.  It is missed in weakly coupled string theory because the Planck scale is infinitely shorter than the string scale in the limit of zero coupling.  This discreteness {\it is} evident in AdS/CFT.  For example, the radius of the spheres in $AdS_n \times S^k$ models is typically determined by some power of an integer.  In $AdS_3$ compactifications admitting a weak coupling expansion, even the string coupling is quantized.  More generally, continuous parameters in AdS/CFT correspond to fixed manifolds of conformal field theories.  These always have many fewer parameters than the compact geometries of the gravitational duals.  The HST theory of AdS models has not yet been worked out, but it will certainly involve a limit of finite dimensional quantum systems which approach the field theory when the proper time becomes of order the AdS radius and the area of causal diamonds goes to infinity. Conformal manifolds arise as a consequence of this limiting process. 

\subsection{Jet states in Finite Causal Diamonds and Energy as an Emergent Conservation Law}

We have demonstrated above how the Hilbert space of Jets arises on the conformal boundary of Minkowski space, by considering representations of the Generalized Super-BMS algebra, which satisfy a set of constraints.  We now want to discuss how this structure emerges as a limit of a theory describing finite area diamonds.  We begin by constructing the matrices 
$M (a\ b ;\ k\ l)_s^t = \psi_{(a\ k) {s\ s_2 \ldots s_{d-2}}} \psi^{\dagger\ (b\ l){t\ s_2 \ldots s_{d-2}}} . $  In the limit $N \rightarrow\infty$, these converge to the algebra of matrix ( in the $k,l$ indices) differential forms\footnote{$a$ and $b$ are spinor indices on the sphere, so their tensor product includes all anti-symmetric tensors.} (with Clifford multiplication) on the sphere.  If we also take the cutoff on the Dirac operator on ${\cal K}$ to infinity, we get the algebra of forms on the full $d - 2 + D$ dimensional holoscreen.  

We consider Hamiltonians which are traces of polynomials in the matrices $M$. The constraints defining the jet Hilbert space at infinity, guarantee that different jets do not interact, and that the jets do not interact with the statistically isotropic sea of soft massless particles.  For Hamiltonians which are single traces of $N \times N$ matrices, an analogous breakup into non-interacting systems is a subspace of the Hilbert space on which many matrix elements vanish so that 
the matrices are block diagonal.  This description of scattering states was first encountered in Matrix Theory\cite{bfsstbreview} .  For matrices constructed as bilinears in our cutoff Dirac spinors, this constraint sets of order $N$ of the spinor variables to zero.  

To understand this counting we note that the spinors, for each value of the spinor index, can be viewed as anti-symmetric $U(N)$ tensors of rank $d - 2$.  Thus, we can associate a small subset of indices with a surface of dimension $d-2$ .  Give the surface the topology of a hypercube. The above construction of matrices from spinor variables corresponds to gluing these surfaces together along a $d - 3$ dimensional face of the small hypercubes and constructing the full surface as a space filling curve consisting of hypercubes of various sizes glued together. 

If we have two blocks of the bilinear matrices $M$ , of sizes $N$ and $K$, with $N \gg K$, the number of elements of the spinor variables which must be set to zero 
in order to make the matrices block diagonal is of order $(N - K) K^{d-3}$.  To see this, consider a typical term in the Hamiltonian
\begin{equation} \psi_{k k_1 \ldots k_{d-3}} \psi^{\dagger\ k_1 \ldots k_{d - 3} \mu} A_{\mu}^k , \end{equation} where the indices labeled by small $k$ run from $1$ to $K$ and $\mu$ runs from $1$ to $N$.  The matrix $A$ is some long string of bilinears in the fermions, with $d-3$ of the bilinear indices contracted together to make an $N \times K$ matrix.   We see that any fermionic variable all of whose indices run from $1$ to $K$ interacts with the majority of the variables only through a fermion with $d - 3$ small indices and one index that runs from $1$ to $N - K$.  Thus, to decouple the small block from the rest, we need only constrain the state so that this set of coupling variables vanish.   In terms of the picture of a small spherical cap, separated from the larger part of the sphere by an annulus, the annulus is taken to have an {\it area} $K^{d-3} N$.   

The probability\footnote{Douglas Stanford has emphasized to me that this is a combined statistical and quantum probability.  That is, we are computing the combined statistical and quantum probability of being in a quantum state such that the indicated components $\psi_{\ldots}$ vanish. The number of variables being constrained $K^{d-3} N$ is much smaller than the total number $N^{d-2}$.  For large $N$, Page's theorem then tells us that the density matrix on the tensor factor of the Hilbert space defined by the constrained variables is close to maximally uncertain.  If each independent variable can take on the same number, $v$, of values, then the probability of finding the constraint satisfied is $v^{- {K^{d-3}N} }$.} of finding such a constrained state, when picking a random state from the Hilbert space, is $e^{- c K^{d-3} N}$ which is the Boltzmann distribution if
we take the energy to scale like $K^{d-3}$ and the temperature $T \sim 1/N$ .  The latter is of course the scaling law for the Hawking temperature of a black hole in any dimension.  The identification of energy with $K^{d-3}$, a generalization of the familiar Matrix Theory rule in $4$ dimensions, is reinforced by three other calculations.   First of all, if we consider incoming states, consisting of blocks of size $K_i$ in a large causal diamond of size $N \gg \sum K_i$, then it is easy to argue, given the scaling of the Hamiltonian 
\begin{equation} \sum_{l = 2}^{l_{max}} a_l N^{ - (2l-2) (d-3)} {\rm Tr}\ (\psi \psi^{\dagger} )^l , \end{equation} which is one power of $N^{d-3}$ smaller than the 't Hooft scaling of this large $N$ tensor model\cite{tensormodels}, that a state with $o(N \sum K_i^{d-3} )$ constraints at time $- N$, must evolve to a state at time $N$, with the same number of constraints to this order in $N$ as $N \rightarrow \infty$.  Thus, $\sum K_i^{d-3} $ is an asymptotic conservation law for this time dependent Hamiltonian\footnote{An extension of this argument can be used to justify the precise power law in the scaling of the Hamiltonian.  A lower power would lead to asymptotic forces that fell off more slowly than Newton's law, and would not admit an S-matrix. A higher power would lead to more asymptotic conservation laws which were powers of the energy.  This is incompatible with Lorentz invariance if there is a non-trivial S-matrix.}. 

Secondly, as $K \rightarrow \infty$ (but always $\ll N$), our variables fill out the full spinor bundle, and we can reconstruct the SUSY algebra.  The energy-momentum in the SUSY algebra is the trace of a bilinear in the fermionic generators, which is proportional to the unit matrix, and so is of order $K^{d-3}$\footnote{The Hamiltonian acting on the $K$ block variables has, in the 
$N \gg K\rightarrow\infty$ limit a term of order $K^{d-3}$ which represents the total energy of the jet, plus terms which fall off with $K$, and represent the splitting and joining of collinear soft gravitons from the jet.}.
Finally, if we take the interaction Hamiltonian to be of the form
above, then it was shown in \cite{holonewton} that the large impact parameter scattering amplitudes for two blocks of size $K_1$, $K_2$ is that expected from a static potential of the form $\frac{K_1^{d-3} K_2^{d-3}}{r^{d - 3}} $.  

Another generic result for the entire class of Hamiltonians described above is the threshold above which particle collisions lead to an equilibrated state with the properties of a black hole.  This occurs if we try to set up a scattering state in which the incoming particle blocks in a causal diamond satisfy $\sum K_i^{d-3} \sim n^{d-3}$, because in such conditions it is impossible to satisfy the constraints defining particle states.    In terms of energies, this is just the statement that 
the Schwarzschild radius of the energy enclosed in the diamond is of order the size of the diamond. For a system obeying the constraints of special relativity, 
the sum of the energies will be replaced by the total center of mass energy.  

Most of our models do not satisfy the relativistic constraint.  Therefore, they would not be compatible with the HST principle that the density matrices for the overlap region of the causal diamonds along two Minkowski geodesics with non-zero relative velocity are unitarily equivalent.  We have not yet found a model satisfying this constraint.  Nonetheless, we do have quantum models with many of the properties one would expect of a theory of quantum gravity.  Each of these models has a scattering matrix with an asymptotically conserved energy.  The spectrum of asymptotic states is covariant under the super-Poincare group.  Large impact parameter scattering scales with energy and impact parameter as one would expect from Newton's law.  The models also have long lived, high entropy meta-stable states, which are formed in particle scattering when the Schwarzschild radius corresponding to the total energy is of order the size of the causal diamond in which the scattering takes place.  These states decay by emission of a thermal spectrum of particles with a temperature inversely proportional to their size.

The ultimate aim of the HST program is to find a version of these models which satisfies the compatibility constraints for Hamiltonians describing physics along any Minkowski geodesic.  It follows from this that the limiting S-matrix will be super-Poincare invariant.  

\section{Conclusions}

The S-matrix formalism is inappropriate for theories of massless particles, in Minkowski space, because there are processes of finite probability, in which an arbitrary number of massless particles of low energy, are emitted in a collision between a small number of incoming particles.  A better description of such systems is in terms of jets of particles, or equivalently, in terms of an algebra of current densities at null infinity.  The conservation laws for these currents at the boundary, tell us that if there are only massless particles at the boundary, then only one component of each current has a constant non-zero limit on the conformal boundary. In models containing massless particles of spin higher than one half, the relevant currents are, in the field theory approximation, limits of currents that are generally only covariantly conserved in the bulk, and the boundary current contains contributions from the relevant gauge particle. 

Among these currents, we always have the stress tensor, whose surviving components at null infinity are the generators of BMS super-translations.  The spectrum of these currents is the dual momentum space to null infinity, and lives on the light cone $P^2 = 0$.  Other generators of the current algebra, which commute with the momentum may be thought of as generalized functions on this null cone.  It appears that measures and half measures are the most singular form of distributions that are necessary.  The algebra of densities is covariant under the Lorentz symmetry of the null cone.  The generator of these transformations is
$P_A \partial_B - P_B \partial_A + \Sigma_{AB} (P) . $  Even in models containing gravity, it does not appear that there is an independent gauge particle associated with the spin density $\Sigma_{AB} (P)$.   The model may also have a larger group of internal symmetry currents $J_{ij} (P)$, transforming in the adjoint representation of a compact Lie group $G$. In addition, there may be operators $Q_{i \alpha} (P)$ which are charged under the center of $Spin (1, d - 1)\times G$.  In non-gravitational quantum field theory, there can be operators non-trivial under the center of $G$, but trivial under $(- 1)^F$. We conjecture, based on empirical evidence from string theory, Matrix Theory, and AdS/CFT, that this is impossible in gravitational models.

The null cone has two branches, connected by the singular point $P = 0$.  Jet representations of the algebra of densities are defined by having a clean separation between the supports of generators at non-zero $P$ and $P = 0$.  For non-zero $P$ the angular support of densities must be restricted to a finite number of spherical caps of finite opening angle.  The zero momentum generators are non-trivial measures on the sphere, whose support vanishes in annuli surrounding the spherical caps.  The Scattering operator maps the jet representation of the density algebra on the past $(P_0 < 0)$ null cone, into that on the future null cone. 

To define these representations carefully, we must view the conformal boundary of
Minkowski space as the limit of the boundary of a finite causal diamond.  This limiting procedure is very different in quantum field theory than it is in models of quantum gravity.  In field theory, the boundary operator algebra is a sub-algebra of the bulk algebra, defined by a set of boundary conditions.  In quantum gravity, there are only boundary algebras.   The ``bulk" algebra of a fixed causal diamond $D$, corresponds to boundary algebras of causal diamonds completely contained in $D$.  The holographic principle states that these algebras are all finite dimensional, for small enough proper time between the tips of a diamond, and that the algebras of smaller causal diamonds are contained in that of $D$ as proper tensor factors.   We organize the collection of diamonds into nested sets corresponding to nested proper time intervals along time-like trajectories. Each time-like trajectory is an independent quantum system.  Causality is enforced along each trajectory by making the Hamiltonian time dependent.  The Hamiltonians describing evolution for some interval of proper time, contains only operators acting in the tensor factor of the full Hilbert space corresponding to that interval.  

These systems are knit together into a single quantum space-time by specifying, for each pair of causal diamonds along two different trajectories, a tensor factor in each of the Hilbert spaces, such that the eigenvalues of the density matrices for this tensor factor, prescribed by the initial conditions and Hamiltonians of two otherwise independent quantum systems, are the same.  This infinite collection of consistency conditions is the quantum analog of the principle of general relativity. 

The variables in this formalism are restrictions of the algebra of densities on the conformal boundary.  Taking the hint from string theory that this algebra is generated by fermionic spinor operators, we guess the form of the restriction to be a cutoff of the expansion of the the fields on the conformal boundary in terms of eigenspinors of the Dirac operator on the holographic screen.  This leads to a generalization of the UV/IR correspondence, familiar from AdS/CFT, but specifies more precisely what kind of UV cutoff corresponds to a particular causal diamond. 

This general approach to models of quantum gravity leads one to a number of very general conclusions. The first is that the geometry of space-time is {\it not} a fluctuating quantum variable.  Quantum avatars of both the causal structure and conformal factor of a background space-time are built into the HST construction of the corresponding quantum theory.  The second is that compact factors of the geometry, with finite volume in Planck units, are fuzzy geometries, which have only discrete parameters.  Those discrete parameters are incorporated in the operator algebra, and are not fluctuating quantum variables: the theory has no continuous moduli.  The cosmological constant in Planck units is also a discrete parameter, determined by the total number of states (if $\Lambda > 0$) or the coefficient in the high temperature entropy (if $\Lambda < 0$).  All of these features are consistent with Jacobson's\cite{ted} view of Einstein's equations as the hydrodynamic equations of systems obeying the Bekenstein-Hawking-
t Hooft - Fischler - Susskind - Bousso relation between area and entropy.
Hydrodynamic equations are only quantized in special situations describing low amplitude low energy fluctuations around the ground state.  They are valid as classical equations even in high entropy regimes where the concept of phonons may fail.

The definition of particle/jet states in HST gives more insight into how effective quantum field theory emerges from the underlying formalism.  It is motivated by the ``jet isolation criterion" on the conformal boundary.  Models of quantum gravity in Minkowski space will always have finite amplitudes in which an arbitrarily large
number of very low energy gravitons are present in the initial and final states.
The jet isolation criterion groups all the energy penetrating the holographic screen at infinity into a finite number of finite energy bearing spherical caps, of finite opening angle, and all of the rest of the energy less than some cutoff $\epsilon$ which can go off at arbitrary angles.  The jets must be isolated from the soft graviton/gauge boson cloud by annuli through which no energy flows.

At null infinity, this isolation criterion requires a delicate mathematical definition, which is not yet available.  HST makes it more precise, by describing the analogous criterion in a finite causal diamond. That description has a number of crucially important properties:

\begin{itemize}

\item The total entropy of a diamond of radius $N$ in Planck units is $N^{d-2}$. Jet states are defined by imposing of order $N$ constraints on these variables\footnote{The real constraint is that the quantum state has an overlap of order one with a state in which the operator constraints are satisfied.}.
For a large class of time dependent Hamiltonians for these systems, the coefficient of $N$ in the asymptotic number of constraints is an asymptotic conservation law, which we interpret as energy.  It follows immediately that 
the probability of finding a state of energy $E$ is thermal, with a temperature of order $N^{-1}$.  This explains both the Hawking temperature of black holes and the temperature of de Sitter space.

\item From the same considerations one concludes that the maximal entropy of states that have an order one probability of having particles is bounded by an inequality of the form $E N \ll N^{d-2}$, where $E$ is the particle energy.  The threshold where this inequality fails is parametrically the same as the geometrical criterion that the Schwarzschild radius of the total particle energy is equal to the radius of the causal diamond.  It follows that all candidate HST models have processes in which particle collisions produce black hole like equilibrium states.

\item  All of our candidate Hamiltonians have the property that, as $N \rightarrow \infty$ the bulk of the degrees of freedom in a causal diamond decouple and carry zero energy.  That is, they behave like gauge modes.  The diffuse cloud of low energy gravitons is continuously connected to these gauge modes.  This is the origin of soft graviton theorems.  If $N$ is held finite and the system evolves forever with the Hamiltonian $H(N_{max})$ then we obtain a system with the properties of stable de Sitter space:  everything involves to a thermalized state with temperature of order $N_{max}^{-1}$. 

\end{itemize}

The aim of this paper has been to show that a general principle of quantum gravity, the fact that all physical degrees of freedom in a causal diamond lie on its boundary, lead to a picture in which the operator algebra in a causal diamond is a subalgebra of the algebra of operators at null infinity.  Most states of quantum field theory in a finite diamond, are not states of a quantum theory of gravity.  The strong holographic principle, according to which the logarithm of the dimension of the Hilbert space associated with a diamond is one quarter of the area of the holographic screen, makes the opposite point: most of the states in the quantum gravity Hilbert space, cannot be associated with states localized in the bulk of the diamond.  Instead, bulk localized states have a probability of order one, that of order $N$ of the degrees of freedom in a diamond of entropy $N^{d-2}$ are frozen.  At null infinity, where $N \rightarrow\infty$, this corresponds to the statement that a jet of energy $E$ is  separated from the zero momentum cloud by an annulus whose area in Planck units is $EN$.  This postulate explains the thermal behavior of both black hole and de Sitter horizons.

We close by noting two general features of our approach, which require further conceptual elaboration\footnote{This phrase is meant to distinguish these problems from that of finding models which satisfy the consistency relations between trajectories related by Lorentz boosts, and which will lead to a Lorentz invariant S matrix. This is a technical problem, but a crucial one which requires 
some new insight.}.  The first is the relation between our approach to the super-BMS algebras as the generators of the algebra of observables on the conformal boundary of Minkowski space, and the more traditional definition of these algebras as {\it asymptotic symmetries}.  The asymptotic symmetry viewpoint has been advanced in the very interesting recent work of Strominger\cite{strom} and his collaborators.  Strominger makes the point that, apart from the total momentum operator, the BMS super-translations should be thought of as spontaneously broken symmetries.   In conventional QFT, spontaneously broken symmetry generators do not act on the Hilbert space of physical states.  However, in a conventional QFT we always define an S-matrix in a Fock space of asymptotic states.  Spontaneously broken symmetry generators transform a Fock space state into a coherent state of Goldstone bosons, which is orthogonal to every vector in the original Fock space.   An intriguing conjecture is that the Hilbert space discussed here, a jet representation of the asymptotic algebra of densities, is the proper venue for the action of spontaneously broken symmetry generators, including the super-BMS algebra.

A second intriguing line of thought is the evolution of the concept of particle in light of the necessary emission of soft radiation.  Particles are replaced by jets, which are quantum systems with an infinite number of states.  The emission of approximately colinear soft radiation acts as a heat bath, which decoheres the variables describing the asymptotic jet trajectory in space-time.  Thus, the conventional description of spreading of single particle wave packets, which measures the regime of validity of the notion of a classical particle trajectory, breaks down when the single particle wave function spreads by more than the opening angle of the jet.  From that point on, one can apply classical statistics, and in particular, {\it Bayes Rule} for conditional probabilities, to the quantum predictions for the probability of finding a particle at a particular place and time.  More precisely, these decoherent probabilities are those for finding the central trajectory of a jet at a particular place and time.

We are used to interpreting the observation of what look like classical particle tracks in detectors in terms of the interaction of the particle with the many degrees of freedom of the detector.  The particle track decoheres because it leaves an imprint on the collective coordinates of regions of the detector which contain many atoms.  The new conjecture is that the interaction of a moving particle with the massless particles it produces, induces a similar decoherence even in empty space.

I do not mean to suggest that the suggestions made in the last few paragraphs are anything more than conjectures.  Whether they are true or not, the main claims of this paper: that scattering of massless particles is best described in a Hilbert space, which is a jet representation of the algebra of densities on the conformal boundary of Minkowski space, and that in models of quantum gravity, the variables in finite causal diamonds are a subalgebra of that density algebra, remain valid.   

\end{document}